%%%%%%%%VERSION OF Aug 11 1996%%%%%%%%%%%%%%%%
\input harvmac
\def\pl#1#2#3{{Phys. Lett. } {\bf B{#1}} (#2) #3}
\def\np#1#2#3{{Nucl. Phys. } {\bf B{#1}} (#2) #3}

\def\prl#1#2#3{{Phys. Rev. Lett. } {\bf {#1}} (#2) #3}

\def\ltap{\ \raise.3ex\hbox{$<$\kern-.75em\lower1ex\hbox{$\sim$}}\ }
\def\gtap{\ \raise.3ex\hbox{$>$\kern-.75em\lower1ex\hbox{$\sim$}}\ }
\def\lc{\Lambda}
\def\ls{\Lambda_{SUSY}}
\def\mt{\tilde M}
\def\sm{$SU(2)\times U(1)$}
\def\smt{$SU(3)\times SU(2)\times U(1)$}
\def\ie{{\it i.e.}}
\def\eg{{\it e.g.}}
\def\CW{{\cal W}}
\def\CK{{\cal K}}
\def\pyidk{PHY-9057135}
\Title{\vbox{
\hbox{BU/HEP-96-21}
\hbox{DOE/ER/40561-270-INT96-00-131}
\hbox{UW/PT-96-10}}}
{\vbox{
\vskip-.6in\centerline{The More Minimal}
\vskip8pt\centerline{Supersymmetric Standard Model}}}
\centerline{A. G. Cohen}
\smallskip
\centerline{\sl Department of Physics, Boston University,
Boston, MA 02215, USA}
\centerline{{\tt cohen@andy.bu.edu}}
\medskip
\centerline{D. B. Kaplan}
\smallskip
\centerline{\sl Institute for Nuclear Theory 1550,
University of Washington}
\centerline{\sl Seattle, WA 98195-1550, USA }
\centerline{{\tt dbkaplan@phys.washington.edu}}
 \medskip
\centerline{A. E. Nelson}
\smallskip
\centerline{\sl Department of Physics 1560, University of
Washington, Seattle, WA 98195-1560, USA}
\centerline{{\tt anelson@phys.washington.edu}}
\noindent
\vskip.3in
\baselineskip 14pt
\noindent
Effective Supersymmetry is presented as a theory of physics
above the electroweak scale which has significant theoretical
advantages over both the standard model and the  Minimal
Supersymmetric Standard Model (MSSM).  The theory is
supersymmetric at short distances but differs significantly
from the MSSM. Flavor symmetry violation is intimately
related to supersymmetry breaking. There is a new physics scale
$\mt\sim$~5--20 TeV which sets the mass of the first two
sparticle families. Supersymmetric sources of CP violation and
flavor changing neutral currents for the first two families
are suppressed. Effective Supersymmetry can be implemented with
automatic suppression of baryon and lepton number violation
and  a dynamically generated $\mu$ term, while maintaining
naturalness in the Higgs sector. There are implications for
new particle searches,  flavor and CP violation experiments,
as well as for the construction of theories of flavor and
dynamical supersymmetry breaking.
\Date{7/96}
%\draftmode

\newsec{Introduction}

Despite the success of the $SU(2)_L \times U(1)_Y$ theory of
electroweak interactions, the mechanism of electroweak
symmetry breaking remains obscure. A multitude of models have
been constructed to explain this symmetry breaking, but none
of them are experimentally confirmed nor theoretically
compelling. The three most popular approaches are the standard
model, the Minimal Supersymmetric Standard Model (MSSM), and
technicolor. Each approach has its virtues as well as its
problems. In this Letter we attempt to systematically combine
the best features of each of these approaches, while avoiding
their respective defects. We employ a  ``bottom up'' line of
reasoning, in which we de-emphasize the role of short
distance dynamics, instead explaining the physics we observe
in terms of  ``accidental'' or ``effective'' symmetries.
Such symmetries are global symmetries obeyed by the lower
dimension operators of an effective field theory as a
consequence of the gauge symmetry and particle content of
that theory, without being symmetries of the fundamental
dynamics.

We are prompted by this reasoning to postulate that the world
possesses such an effective supersymmetry at scales below
1~TeV, which explains how the electroweak hierarchy can be
stable; the meaning of effective supersymmetry will be made
more precise below. The theory which we construct (and which we
call Effective Supersymmetry) is quite unlike the MSSM, since
we require it to possess additional effective symmetries which
help to explain the absence of flavor changing neutral
currents (FCNC), lepton flavor violation, baryon (B) and
lepton (L) number violation, and electric dipole moments
(EDMs) of the neutron and electron. Effective Supersymmetry
has distinctive theoretical and experimental  consequences: 
there are new interactions  for quarks and leptons
characterized by an undetermined scale $\Lambda$ and
supersymmetry breaks at a scale $\ls\equiv\sqrt{F_s} \le
\Lambda$. The scale $\mt\equiv F_s/\Lambda$ is fixed to be
5--20~TeV. It is the existence of these new scales above $M_W$
which makes an analysis of effective symmetries both
nontrivial and fruitful. Between the scales $\Lambda$ and
$\mt$ the particle content of the effective theory is that of
the MSSM, with SUSY spontaneously broken. However most of the
scalar fields which  would be present in the MSSM have a mass
of size $\mt$---in particular, all of the sparticles of the
first two generations. The SUSY partners with masses below a
TeV consist of the gluinos, charginos, neutralinos and third
family squarks and sleptons. 

The line of reasoning leading to Effective Supersymmetry
begins with an examination of the strong and weak points of
the three basic electroweak symmetry breaking models.

\subsec{The Standard Model}

There are no compelling discrepancies between the standard
model and experiment. Furthermore no {\it ad hoc\/} global
symmetries are required to explain the absence of proton
decay, lepton number violation, and the observed pattern of
extremely small FCNC---all of these features result from
accidental symmetries of the renormalizable operators allowed
by gauge invariance. This is a very beautiful feature of the
standard model, but as an effective theory
one must assume that  non-renormalizable operators (which do
not enjoy the accidental symmetries) are greatly suppressed
or absent from the theory.

The standard model has another remarkable feature:
$\Tr\ Y=0$, where $Y$ is the generator of hypercharge on the
left handed fermions. This allows charge quantization to be
explained via gauge unification into a non-Abelian symmetry
\ref\unif{J.C. Pati  and A. Salam, Phys. Rev. {\bf D8} (1973)
1240;  H. Georgi and S. Glashow, \prl{32}{1974}{438}.}.

Major deficiencies of the standard model include its lack of
explanations for the absence of strong  CP violation,
and the tremendous hierarchy between the weak and Planck
scales.  Also unexplained are the patterns of lepton and quark
masses and mixing angles built into the Yukawa interactions;
the Yukawa couplings have been regarded as ``flavor spurions''
parameterizing the low energy effects of the breaking of chiral
flavor symmetries at short distance (see for example
\ref\fn{C.D. Frogatt and H.B. Nielsen, \np{147}{1979}{277}.}).
Each of these deficiencies may imply new physics
at scales (perhaps much) higher than the electroweak scale.
However the most serious criticism of the  standard model is
that it cannot be valid above a scale of about $1$~TeV, without
an unnatural cancellation between short and long distance
contributions to the Higgs mass \ref\thooft{G. 't Hooft, 1979
Cargese Lectures, published in Recent Developments In Gauge
Theories, Proceedings, NATO Advanced Study Institute New York,
USA: Plenum (1980).}.

A fundamental dichotomy haunts modifications of the standard
model: new particles and interactions at the TeV scale which
cure the naturalness problem risk destroying the accidental
symmetries of the standard model which are so successful in
explaining phenomenology. 

\subsec{The Minimal Supersymmetric Standard Model (MSSM)}

The MSSM \ref\mssmhiggs{See {\it Review of Particle
Properties}, Phys. Rev. {\bf D54}, 1996, pp. 687-692 and
references therein.} has 122 free parameters \ref\savas{see,
{\it e.g.}  S. Dimopoulos and D. Sutter, \np{452}{1995}{496}.}
which, not surprisingly, can be chosen to agree with
experiment. However the model provides no explanation for
experimentalists' failure to observe B or L violation, large
FCNC, or EDMs---these features arise by a conspiratorial
adjustment of parameters, often with an appeal to new exact
and approximate global symmetries. It does not address the
origin of the scale of weak symmetry breaking, nor the scale
of soft supersymmetry (SUSY) breaking. Its best feature is
that it is natural up to extremely short
distances\nref\dimgeorg{  H. Georgi and S. Dimopoulos, Nucl.
Phys. {\bf B193} (1981) 150; N. Sakai, Z. Phys. {\bf C11}
(1981) 153.}\nref\noquad{L. Girardello and M. Grisaru, Nucl.
Phys. {\bf B194} (1982) 65.} \refs{ \dimgeorg, \noquad}: with
any cutoff below $M_{pl}$ all quantum corrections to MSSM
parameters are smaller than the parameters themselves. 

\subsec{Dynamical Electroweak Symmetry Breaking}

Technicolor \ref\weinsuss{S. Weinberg,  Phys. Rev. {\bf D13}
(1976) 974, {\it  ibid} {\bf D19} (1979) 1277;  L. Susskind,
Phys. Rev. {\bf D20} (1979) 2619; for a review see E. Farhi and
L. Susskind, Phys. Rept. {\bf 74} (1981) 277.} aims to
eliminate the need for unnatural  scalars below $M_{pl}$. This
model takes as its inspiration QCD---which naturally explains
the small ratio $\Lambda_{QCD}/M_{pl}$ in terms of
nonperturbative dynamics---and postulates that the small
number $M_W/M_{pl}$ is similarly generated nonperturbatively
by new strong interactions.  However it is not apparent how to
construct an experimentally acceptable renormalizable model
without re-introducing scalars. Thus viable models tend to
suffer from the disease they sought to cure.

\newsec{Effective Supersymmetry}

We desire a theory that incorporates all of the positive
attributes listed above:  nonperturbative generation of the
electroweak hierarchy, unifiable weak hypercharge, accidental
B and L symmetries, suppression of FCNC and CP violation, and
agreement with experiment. We now present the details of our
argument that  a minimal extension of the standard model that
incorporates these successful features is Effective
Supersymmetry.

\subsec{The Observed Spectrum and Naturalness}

We begin by adopting fundamental scalar fields and Yukawa
interactions which have the virtue of being successful at
reproducing the intricate pattern of  masses and mixing angles
observed in Nature, and which can be made consistent with
precision electroweak measurements. Once light scalar fields
are admitted to the theory, the large hierarchy between
the weak and Planck scales can be stable against radiative
corrections if one embraces supersymmetry as well.

It is important for our subsequent discussion to address the
question of how much supersymmetry is enough to maintain
naturalness.  As was pointed out when the MSSM was introduced
in \dimgeorg, exact supersymmetry is not absolutely
necessary: a theory with soft SUSY violating operators at the
weak scale is sufficient for maintaining naturalness up to the
GUT scale. However, from a low energy ({\it e.g.} effective
field theory) point of view---for which the GUT scale is
irrelevant---naturalness can be maintained  even with hard
(dimension 4) SUSY violation up to a scale significantly
higher than the ``'t Hooft scale'', $\sim 1$~TeV. Such an
effective theory must derive from a more fundamental theory
which is supersymmetric at high energies, and so the effective
theory can be thought of as the result of integrating out
heavy particles from a softly broken supersymmetric theory. 

In an attempt to raise the naturalness scale above 1 TeV, the
first problem one encounters is the possibility of a 
tree-level Fayet-Iliopoulos term \ref\fayetill{P. Fayet and J.
Iliopoulos, Phys. Lett. {\bf B51} (1974) 461.} for weak
hypercharge, which contributes directly to the Higgs mass
squared \foot{The importance of the Fayet-Iliopoulos term was
emphasized in \ref\dimgiu{S. Dimopoulos and G. Giudice,
\pl{357}{1995}{573}. }, in the context of naturalness of SUSY
GUTS.}.  To prevent such a term with a natural coefficient of
$M_{pl}^2$ in the full theory, we deduce that 
\eqn\yfull{\Tr\ Y=0\ ,}
where the trace is over all particles below the Planck scale.
Thus this nontrivial constraint (which is satisfied by the
standard model) is not only required for the gauge symmetry
to be unifiable, but is also a prerequisite for maintaining
naturalness.  Preventing such a term in the full theory is not
sufficient, however, since integrating out heavy particles can
induce a (finite) $U(1)_Y$ Fayet-Iliopoulos term proportional
to $(\alpha_1/4\pi)\Tr\ Y M^2_h$, where $M^2_h$ is the mass
squared matrix of any heavy scalars one has integrated out. 
Assuming that the scale $M_h$  is much greater than 1~TeV, it
follows that naturalness restricts the heavy spectrum to
satisfy
\eqn\mhconst{\Tr (Y \,M_h^2) \simeq 0\ ;}
this constraint can be satisfied if the heavy particles
transform as multiplets  of a non-Abelian global symmetry that
contains $Y$ (such as $SU(5)$), or if $M_h$ is proportional to
some charge $Q$ which has no $Q^2Y$ anomaly, such as
$Q=(B-L)$.  Given eq. \yfull, it follows that the low  energy
spectrum will also exhibit traceless hypercharge; eq.
\mhconst\ then prohibits generation of a Fayet-Iliopoulos term
in the effective theory proportional to  $(\alpha_1/4\pi)
M^2_h$ (where $M_h$ serves as the cutoff of the effective
theory).

In order to raise the naturalness scale of the effective
theory above the 't Hooft scale,
quadratic divergences must cancel at least to order $\alpha/4\pi$ in
the effective theory.  This is tantamount to requiring that all
one-loop quadratic divergences cancel in the limit that the
Yukawa couplings of all the quarks and leptons, with the
exception of the top, are set to zero. (An exception occurs
in the large $\tan\beta$ regime with two light Higgs doublets,
in which case the bottom Yukawa coupling must be retained as well).
This is possible so long as the  spectrum below $\sim
1$~TeV includes left- and right-handed top squarks, a left
handed bottom squark, Higgsinos, a bino and a wino, all with
dimension 4 interactions as given by SUSY, up to order
$\alpha/4\pi$ corrections.  Although not required for
one-loop naturalness, theories with light winos and binos
typically have a light gluino as well.  Note that both the up-
and down-type Higgsinos are required to maintain gauge
invariance in the effective theory due to the triangle
anomaly, even if there is only a single light scalar Higgs doublet.

We  conclude that the above spectrum provides a minimal
effective supersymmetry at low energy which eliminates
one-loop quadratic corrections to the Higgs mass squared.
Two-loop graphs give 20~TeV for the scale where naturalness
would break down without a supersymmetric spectrum. The scale
$\tilde M$ of the first and second generation sparticles must
therefore satisfy\foot{This bound assumes that contributions
to the Higgs mass from 2-loop diagrams computed with a cutoff
$\mt$ should be $\ltap M_Z$. Our bound is less stringent than
the 2--5~TeV bound given in ref. \dimgiu\   because our
bottom-up approach does not assume the SUSY breaking
parameters are generated at the Planck scale, and so does not
include the effects of renormalization group running from
$M_{GUT}$ .}
\eqn\natural{\tilde M\ltap \hbox{20 TeV}\ .}

\subsec{The electroweak hierarchy}

While supersymmetry stabilizes the electroweak hierarchy, it
does not explain it. Technicolor offers the most compelling
explanation for this hierarchy, namely that it arises from
dynamical symmetry breaking. Therefore we extend the gauge
symmetries of effective supersymmetry to $G\times SU(3)\times
SU(2)\times U(1)$, where $G$ is a gauge group containing  new
strong interactions---which we call ``superglue''---which
generates the observed hierarchy nonperturbatively. The
agreement between precision electroweak experiments and
standard model calculations implies that the scale of these
new superglue interactions must be well above $M_W$, and that
superglue interactions decouple. Because of the necessary
relation between the scale of weak symmetry breaking and SUSY
breaking, we postulate that superglue is responsible for 
breaking both symmetries: the potential for the Higgs scalar
is determined by supersymmetry breaking terms. Note that the
SUSY breaking and electroweak breaking scales, although
related,  may be proportional to different powers of the 
superglue scale, such as  occurs in SUSY breaking models where
non-renormalizable terms in the superpotential are responsible
for supersymmetry breaking (see, {\it e.g.}  \ref\iss{K.
Intriligator, N. Seiberg and S. Shenker,  Phys. Lett. {\bf
B342} (1995) 152.}). 

An important consequence of enlarging the standard model gauge
group is that in general there are new accidental symmetries. 
Thus the symmetry group above $\lc$ is $G\times SU(3)\times
SU(2)\times U(1)\times\CA$, where $\CA$ is the group of such
accidental symmetries;  the imposition of $G$ gauge symmetry
can make $\CA$ much larger than the accidental symmetry group
of the MSSM and can account for a number of the standard
model's successes which are quite mysterious in the MSSM. 

\subsec{Flavor Changing Neutral Current Constraints}

In supersymmetric theories with new interactions for the
first two families at a scale $\Lambda$, the existence of
non-renormalizable FCNC operators suppressed by powers of
$\Lambda$ typically constrain this scale  to be larger  than
several hundred TeV. However the constraints become much more
severe with softly broken SUSY, which allows FCNC through {\it
super}-renormalizable interactions in the form of squark
masses and trilinear squark couplings. In the absence of any
compelling model of flavor that suppresses these dangerous
interactions, a straightforward explanation for why FCNC are
not observed is that the squarks and sleptons which mediate
FCNC are heavy and have decoupled from low energy physics
\ref\dks{M. Dine, A. Kagan and S. Samuel, Phys. Lett. {\bf
B243} (1990) 250.}.  The large approximate flavor symmetry we
observe in the world then becomes an accidental symmetry  as
it is in the standard model, rather than the result of
conspiratorial short distance physics. 

To suppress CP conserving FCNC without any universality
\dimgeorg\ or alignment  \ref\nirsei{Y. Nir and N. Seiberg,
Phys. Lett. {\bf B309} (1993) 337,  hep-ph/9304307 .} in this
way requires the first and second family squarks and sleptons
to have masses of size $\tilde M$  satisfying \nref\pomarol{A.
Pomarol and D. Thommasini, hep-ph/9507462, to appear in Nucl.
Phys. B.}\nref\ggms{F. Gabbiani, E. Gabrielli, A. Masiero and
L. Silvestrini, ROM2F/96/21,
hep-ph/9604387.}\refs{\dks,\pomarol,\ggms}  
\eqn\fcncbound{\tilde M \gtap \hbox{50 TeV .}}
If all CP violating phases are maximally large, suppression of
CP violating $\Delta S=2$ interactions imposes a stronger
bound\foot{The bounds (2.4) and (2.5) are approximate. 
There are $\CO(1)$
uncertainties due to unknown short distance physics 
and to long distance QCD.}, 
\eqn\cpbound{\tilde M \gtap \hbox{600 TeV .}}

On the other hand, naturalness (eq. \natural) requires the
first two families of squarks and sleptons to be lighter than
$\sim 20$~TeV and constrains the remaining spectrum below
$\sim 1$~TeV  as described in \S{\it 2.1}. 

With first two family squark masses of $\sim 20$~TeV, and with
the mild assumptions of squark university or alignment at the
20\% level and CP violating phases of $\CO(0.1)$, it is
possible to  satisfy the FCNC and CP constraints. Specific
models of the flavor hierarchy may produce more universality
or alignment, and restrict new CP violation as well
\ref\ratt{see, \eg, Y. Nir and  R. Rattazzi, RU-96-11,
hep-ph/9603233.}, for instance in the model of ref. \pomarol\
FCNC are adequately suppressed if the first two family squarks
have 5 TeV masses.

We conclude that with top, left handed bottom squark,
chargino, and neutralino masses below $1$~TeV, and the first
two family squark and slepton masses in the range $\tilde M =
\hbox{5--20 TeV}$ \foot{In many clever models with  highly
restrictive global symmetries, even lighter squark masses are
acceptable. We disregard this possibility since such
symmetries appear contrived.}, we are able to combine two of
the best features of the MSSM and standard model respectively:
 naturalness of the electroweak symmetry breaking scale, and
suppression of FCNC. A nontrivial constraint on the heavy
spectrum of the theory is eq. \mhconst, which can result from
either the accidental symmetry group $\CA$, or from a coupling
between sparticles and SUSY breaking dynamics which is
proportional to a charge $Q$ without a $Q^2Y$ anomaly. We show
below that  this novel spectrum can be achieved by having the
first two families of squarks and sleptons couple directly to
the SUSY breaking dynamics, while the third family does not. 

\subsec{CP Violation}

Effective Supersymmetry also cures the SUSY CP problem. The
MSSM with universal soft  masses has two CP violating phases
which are strongly  constrained by the absence of observable
EDMs \ref\dgh{M. Dugan, B. Grinstein and L. Hall,  Nucl. Phys.
{\bf B255} (1985) 413; J. Ellis, S. Ferrara, and  D.V.
Nanopoulos, Phys. Lett. {\bf B114} (1982) 231; J. Polchinski and
M.B. Wise, Phys. Lett. {\bf B125} (1983) 393; W. Buchmuller
and D. Wyler,  Phys. Lett. {\bf B121} (1983) 321; F. del
Aguila, M.B. Gavela, J.A. Grifols and A. Mendez, Phys. Lett.
{\bf B126} (1983) 71.};  with  non-universal masses the
number of independent CP violating phases increases to $43$
\savas. However if the first two families of sparticles
are heavier than $\sim 10 m_{\tilde g}$ (where $m_{\tilde g}$
is the gluino mass) while third family squarks are heavier
than $\sim 550$~GeV \refs{\dks,\pomarol},
then none of these phases lead to unnacceptable
EDMs for the electron or the neutron, even if CP violation is
maximal. On the other hand this may allow
detectable CP violation  at the $B$ factory or in top
production. We discuss this further at the end of this Letter
where we explore experimental implications of Effective
Supersymmetry.

\subsec{B violation}

In the standard model, the observed stability of nucleons is
explained by an  accidental baryon number symmetry. Such an
explanation is lacking in the MSSM where disastrous  dimension
4 and 5 B violating operators have to be excluded by the
imposition of global symmetries. The dimension 5 operators are
particularly troublesome as they have to be suppressed by a
scale greater than $M_{pl}$ to be phenomenologically
acceptable.

If we adopt the standard model solution, and B conservation
arises as an accidental symmetry, we must assume that the new
gauge group $G$ forbids dimension 4 and 5 B violating
operators in the full theory. Since at least two generations
are necessary to construct any dimension 4 or 5 B violating
operator, this suggests that at least the first two
generations of quark superfields carry $G$ charges. The new
gauge sector $G$ includes the superglue interactions
responsible for $SU(2)\times U(1)$ and SUSY breaking, but may
include additional gauge interactions.  In one realization of
Effective Supersymmetry, $G$ is just the superglue group, and
the first two quark and lepton generations are composites of
constituents that carry superglue (explaining why they couple
more strongly to SUSY breaking than the third family). 
Another possibility is that there is a spontaneously broken
gauge interaction---either an Abelian or non-Abelian gauged
flavor symmetry---which communicates SUSY breaking in the
superglue sector to the quarks and leptons, and that this
messenger interaction forbids low dimension B violating
operators. In the next section we outline examples of these
two realizations of Effective Supersymmetry.  In both cases we
obtain the squark  spectrum advocated above, and suppression
of B violation can be automatic. 

\subsec{L violation}

The observed conservation of $e$, $\mu$, and $\tau$ lepton
numbers is well explained in the standard model, where the
lepton flavor symmetries  are an accidental  consequence of
the gauge structure of the theory. In contrast, the  gauge
symmetries of the MSSM allow dimension 2, 3, and 4 lepton
violating operators in the form of misaligned slepton mass
matrices, as well as superpotential terms of the form $LH_u$,
$Q\bar D L$ and $LL\bar E$ ($L$, $\bar E$, $Q$, $\bar D$ and
$H_u$ are the lepton doublet, conjugate electron, quark
doublet, conjugate down-type quark, and up-type Higgs
superfields respectively).  These last three operators violate
overall lepton number and can contribute to neutrino masses,
while all of the operators can contribute to $\mu\to e\gamma$
and $\mu\to 3e$.  As in the case of FCNC discussed above,
lepton violation in the slepton mass matrices is sufficiently
suppressed if the first two generations of sleptons have
masses $\gtap 4$~TeV \ggms.  The absence of the dimension 3
and 4 superpotential operators above the scale $\Lambda$ can be
understood once again if lepton flavor symmetries are part of
the  accidental symmetry group $\CA$ of the full theory. This
requires that an appropriate combination of the  $L$, $E$ and
$H_u$ fields transform under the group $G$.

\subsec{\sm\ breaking}

In addition to third generation squarks, gauginos and
Higgsinos, the low energy Effective Supersymmetry spectrum
must include at least  one light scalar Higgs  doublet, but  not
necessarily both doublets of the MSSM.  In one  composite
realization of Effective Supersymmetry described in the next
section,  the $H_d$ doublet has a mass of the same scale, $\mt$,
as the heavy squarks.  This leads to a naturally large value
for $\tan\beta$,  of size $\sim\mt/M_W\sim\CO(100)$.

\subsec{Summary}

We conclude that Effective Supersymmetry can realize the best
features of the standard model, the MSSM, and technicolor and
its variants, with the following features: 
\item{1.} The world is supersymmetric above  $\sim 20$~TeV;
\item{2.} A new gauge group $G$ exists which contains a
strongly interacting ``superglue'' sector that
nonperturbatively generates the SUSY symmetry breaking scale
$\sqrt{F_s}$ as well as the \sm\ symmetry breaking scale. The
only constraint on $\Lambda$ and $F_s$ is $\tilde M\sim
\hbox{5--20 TeV}$, where $\tilde M \equiv F_s/\Lambda$; 
\item{3.} Above the scale $\Lambda$ matter fields (or their
constituents, in a composite model) carry $G$ interactions which
forbid renormalizable  B  and  L  violating operators, as well as
dimension 5  B  violation.
\item{4.} The first two generations couple more strongly to
SUSY breaking than the third,  and the respective squarks and
sleptons are heavy, with masses at the scale $\tilde M$; 
\item{5.} The top squarks and left-handed bottom squarks are
much lighter, with masses $\ltap 1$~TeV. 
\item{6.} The weak gauginos and higgsinos also have masses
$\ltap$ 1 TeV;  
\item{7.} Naturalness allows the gluino to be heavier than 1
TeV. However if we assume that the gluino is an elementary
particle which is weakly coupled at high energies, then it can
not be strongly coupled to SUSY breaking and will also be
lighter than $\sim \hbox{1 TeV}$. 
\item{8.} Only one linear combination of the two Higgs
doublets of the MSSM need  appear in the effective theory
below $\mt$, while the upper bound on the lightest Higgs
scalar is $\approx 120$~GeV (as in the MSSM). 
\item{9.} The tau sleptons and right-handed bottom squark
masses may be light, or as heavy as $\mt$; however the
constraint \mhconst\ must be satisfied.

\noindent
With these features, Effective Supersymmetry allows a natural
gauge hierarchy, while succeeding where the MSSM fails: namely
by simultaneously explaining how the world can be
supersymmetric at high energies while looking so much like the
standard model at low energies. Effective Supersymmetry 
ameliorates the MSSM's serious problems with FCNC and
excessive weak  CP violation without assuming universality in
the squark sector.  Furthermore, it provides a simple
framework for understanding the suppression of  B and  L 
violation.  In the next section we describe two different
realizations of Effective Supersymmetry.

\newsec{Realizations of Effective Supersymmetry}

There are at least two distinct ways to implement Effective
Supersymmetry: either the first two generations are composite
with constituents that carry superglue, or else the matter
fields are fundamental and communicate with the superglue
sector through some gauged flavor symmetry with the first two
generations coupling more strongly to SUSY breaking than the
third.  We now sketch these two realizations.

\subsec{Effective Supersymmetry with composite quarks and leptons}

The minimal extended gauge interaction necessary to implement
Effective Supersymmetry is superglue alone.  Then, as
discussed above,  B  can arise as an accidental symmetry above
$\Lambda$ if the first two generations of squarks and sleptons
are composed of constituents carrying superglue; accidental L
conservation requires a more model dependent charge
assignment. The effective theory below the scale $\Lambda$
contains the superfields of the MSSM as well as superfields
responsible for SUSY breaking, and operators of higher
dimension suppressed by powers of $\Lambda$.  The effective
K\"ahler potential arising from nonperturbative dynamics is
generic, containing all operators allowed by symmetry.  In
contrast the effective superpotential is known to be
non-generic \ref\gen{N. Seiberg, Phys. Lett. {\bf B318} (1993)
469.}. In order to suppress Fayet-Iliopoulos terms, we will
assume the accidental approximate  symmetry group $\CA$
contains  a non-Abelian factor with hypercharge as a subgroup,
which implies eq. \mhconst. Thus $\Tr\ Y$ must vanish separately
on both the fundamental and composite particle sectors.

In the following discussion we demonstrate that Effective
Supersymmetry can be realized if we assume that the gauge
fields, the top and left-handed bottom superfields, and the
up-type Higgs are neutral under superglue; the remaining
particles in the effective theory below $\Lambda$ (those of
the MSSM) are composites of preons which carry superglue.
(There are at least six possibilities for which of the Higgs
and third family particles are elementary which are consistent
with naturalness and eq. \mhconst; \eg\ the $\bar \tau$ and
$H_d$ superfields could also be elementary). In addition we
assume that supersymmetry breaking has a weakly coupled,
O'Raifeartaigh-like description in the  infrared, as occurs in
some dynamical SUSY breaking models \nref\orm{L.
O'Raifeartaigh, Nucl. Phys. {\bf B96} (1975) 331}
\nref\otheriss{P. Pouliot, Phys. Lett. {\bf B367} (1996) 151,
hep-th/9510148; P. Pouliot and M.J. Strassler, Phys. Lett.
{\bf B375} (1996) 175, hep-th/9602031; T. Kawano, YITP-96-5,
hep-th/9602035; E. Poppitz, Y. Shadmi and S. P. Trivedi,
EFI-96-15, hep-th/9605113; EFI-96-24, hep-th/9606184;  K-I.
Izawa and T. Yanagida, Prog. Theor. Phys. {\bf 95} (1996) 829,
hep-th/9602180; K. Intrilligator and S. Thomas, SLAC-PUB-7041,
hep-th/9603158; C. Csaki, L. Randall and W. Skiba,
MIT-CTP-2532, hep-th/9605108;  C-L. Chou, hep-th/9605119; C.
Csaki, L. Randall, W. Skiba and R. G. Leigh, MIT-CTP-2543,
hep-th/9607021.}\refs{\iss,\orm,\otheriss}. 

The minimal realization involves a single composite  chiral
superfield $S$ with effective super- and K\"ahler potentials 
 \eqn\wkpot{W_S=\ls^2 S\ , \qquad\qquad K_S = S
S^\dagger+ {a_1\over\lc^2}S^3S^\dagger+{\rm h.c.}  +{a_2\over
\lc^2}S^2S^{\dagger2}+ \ldots,}
where the $\{a_i\}$ are $\CO(1)$ coefficients parameterizing
unknown strong interaction effects.  Note that $\ls/\lc$ is a
model dependent parameter which varies from $\CO(1)$ to
exponentially small in explicit examples; we treat it as an
unknown. The theory breaks supersymmetry with $\langle
F_s\rangle=\Lambda_{SUSY}^2$, while the scalar components of
$S$ get masses of order $\mt$.

The most general K\"ahler potential for the matter fields in the
effective theory below  $\Lambda$ is constructed by the following
power counting:
\eqn\kpot{K_0+\Lambda^2
{\CK_I}\left({c_i\over\Lambda},{\lambda_i\over 4
\pi}{f_i\over \Lambda}\right)\ ,}
where $K_0$ is the conventional renormalizable kinetic
term, and $\CK_I$ contains all non-renormalizable interactions. The
$\{c_i\}, \{f_i\}$ are the composite and fundamental superfields (and
their conjugates),
while the $\{\lambda_i\}$ are dimensionless couplings between the
fundamental fields and the preons in the theory above $\Lambda$.
In addition there may be small spurions
associated with approximate symmetries of the preon theory.

We may expand the K\"ahler potential  interactions coupling
$S$ and matter fields in powers of $S$ as
\eqn\keff{K_{eff} = \left({S\over \Lambda} +
h.c\right) K^{(1)}+ {S^*S\over \Lambda^2} K^{(2)}+\ldots\ ,}
where the $K^{(i)}$ are functions of superfields of  the form
given in eq. \kpot. Below the scale $F_s$, $K^{(1)}$ 
contributes  terms that can be written in supersymmetric form
as an effective superpotential (as well as other operators);
$K^{(2)}$ contributes to SUSY violating scalar interactions
(masses, trilinear couplings, as well as ``hard'' SUSY
violating couplings). 

Interactions between $S$ and standard model gauge fields are of the form
\eqn\sgauge{\left[n_i {\alpha_i\over 4\pi}{S\over \Lambda}
\CW_i\CW_i\right]_F\ ,}
where $n_i$ is a numerical factor proportional to the index of the
gauge charge in the preon theory above the scale $\Lambda$; one can
easily imagine that the $n_i$ are as large as $\CO(10)$.

Several salient features of the effective theory below
$\Lambda_{SUSY}$ arising from the operators \keff\ and  \sgauge\ are:
\item{1.} \smt\ gaugino masses that  arise from the operator \sgauge\
 are of size \eqn\gino{m_i=n_i(\alpha_i/4\pi) \tilde M\ ,}
 with $\tilde M
\equiv F_s/\Lambda$.

\item{2.} $LL$ and $RR$ squark and slepton mass matrices for
the first two generations come from a term in $K^{(2)}$ of the
form $z_{ij} \Phi^*_i\Phi_j$ (where $z_{ij}=\CO(1)$)   which
yields 
\eqn\squarkmi{\tilde m_{ij} \sim z_{ij} \tilde M\ .}

\item{3.} Third generation $LL$ and $RR$  squark and slepton
masses also arise from $K^{(2)}$ but are suppressed by
perturbative couplings to the constituents of $S$---denoted
here $\lambda_3$---and are given by $\sim {\lambda_3 / 4\pi}
\mt$. For $\lambda_3\sim 1$, the same size as the top quark
Yukawa coupling, third family sparticles have masses over an
order of magnitude less than their counterparts from the first
two families. 

\item{4.} SUSY breaking Higgs masses arise from $K^{(2)}$ as well.
As $H_u$ is fundamental while $H_d$ is composite, their masses are
given by
\eqn\hmass{m_{H_d} \sim  \tilde M\ , \qquad m_{H_u}\sim
{\lambda_H\over 4\pi} \tilde M\ ,}
where $\lambda_H$ parameterizes the coupling of $H_u$ to the
constituents of $S$.  It follows that there is a single Higgs in the
low energy theory:
\eqn\lhiggs{H = \sin\beta\, H_u + \cos\beta\, H_d^\dagger}
with
\eqn\tbeta{\tan\beta \sim {4\pi\over \lambda_H}.}

\item{5.} The ``$\mu$ term''(which contributes to Higgsino
masses) comes from $K^{(1)}$---an example of the
Giudice-Masiero mechanism \ref\gmmech{G.F. Giudice and A.
Masiero, Phys. Lett. {\bf B206} (1988) 480.}---and is the
same size as $m_{H_u}$ given in eq. \hmass. 

\item{6.} The scalar $H_u H_d$
mass term (the ``$B\mu$ term'') comes from
$K^{(2)}$ with size $\sim {\lambda_H/4 \pi} \mt^2$.

\item{7.} Yukawa interactions can arise from both $K^{(1)}$ and
the non-generic superpotential $W$. However the $b$ quark Yukawa
coupling is $\CO(1)$ and must come from $W$.

\item{8.} SUSY violating trilinear scalar couplings  can come
from both $K^{(1)}$ and $K^{(2)}$, with maximum size
$\mt\mu/\Lambda$. (They may be further suppressed by factors
of $\lambda/ 4\pi$ or spurions). Note that these will be quite
small for elementary scalar fields such as the top squark. For
$\Lambda\gg\mt$, nonsupersymmetric trilinear terms are
suppressed for all scalars, avoiding any problems with vacuum
stability \ref\casas{  J.A. Casas and S. Dimopoulos,
CERN-TH-96-116, hep-ph/9606237.}. 

\item{9.} The fermion partner of $S$ becomes  the massless
Goldstino $G$;  it serves as the longitudinal modes of  the
gravitino, which acquires a mass 
\eqn\mthreehalf{m_{3/2}={F_s \sqrt{8\pi}\over\sqrt3 M_{pl} }\  , }
and has couplings proportional to $1/F_s$.

\noindent
This theory is a successful realization of Effective
Supersymmetry provided that $\tilde M = F_s/\Lambda\sim
\hbox{5--20 TeV}$ (so that FCNC and CP violation are
suppressed), $\lambda_H/4\pi \sim 10^{-2}$ (to provide the
correct electroweak scale), and $\lambda_3/4\pi \sim 10^{-1}$
(to ensure a light enough stop). Note that  eqs. \lhiggs,
\tbeta\ imply $\tan\beta=\CO(100)$  and  the light Higgs is
mostly $H_u$, which involves no fine tuning in this theory.

\subsec{Effective Supersymmetry with gauged flavor interactions}

As another  realization of Effective Supersymmetry we can
consider a theory above the scale $\Lambda$ in which the
ordinary quarks and leptons carry only new weak gauge charges.
These new weak interactions would then be responsible for
communicating with the strongly interacting sector which
breaks supersymmetry. As discussed above, to fully implement
Effective Supersymmetry these interactions must serve double
duty: they must forbid B and L violating renormalizable
operators and they must distinguish the coupling of the first
two generations to the SUSY breaking sector in a way which
keeps the third family squarks light, while allowing the first
two family squarks and sleptons to become heavy.

As one concrete example, we may consider a new Abelian gauge
symmetry which appears anomalous (with a Green-Schwarz anomaly
cancellation mechanism operating near the Planck scale
\ref\greenschwartz{M. Green and J. Schwarz, Phys. Lett. {\bf
B149} (1984) 117.}) with only the first two families carrying
a non-zero value of this charge. The first two families of
squarks and sleptons would then get tree level masses from
this $U(1)$ gauge interaction, while the third family of
squarks and sleptons would only receive masses at higher loop
order or via supergravity, and would thus be naturally lighter.
(Anomalous $U(1)$s which communicate non-perturbative SUSY
breaking to the ordinary quarks and leptons in this way have
recently been introduced in \ref\dude{P. Binetruy and E.
Dudas, LPTHE-ORSAY-96-60, hep-th/9607172.})\foot{After
completion of this work ref.~\ref\dvalipom{G. Dvali and A.
Pomarol, CERN-TH/96-192, hep-ph/9607383.} appeared, which
overlaps with some of the ideas in this section.}. In these
models squarks receive comparable mass contributions from
supergravity and this $U(1)$ interaction. A suitable
re-adjustment of the scales and $U(1)$ charges involved could
produce the Effective Supersymmetry spectrum. 

\newsec{Implications of Effective Supersymmetry}

\subsec{Flavor}

The large value of $\tan\beta$ obtainable in Effective
Supersymmetry can explain the small $m_b/m_t$ ratio without
fine tuning. Furthermore, since the particles of the first two
families carry gauge interactions different from the top, it
is natural to try and relate this to an explanation for the
lightness of these fermions. It is possible for the Yukawa
couplings of the first or first two families to be generated
radiatively and derive only from the K\"ahler potential terms
in eq. \keff. K\"ahler potential terms can give a contribution
to effective fermion-Higgs couplings of order
\eqn\fermionmass{\lambda_{\hbox{eff}} \sim {\mu\over
\Lambda}\le 10^{-2}\ .} 
This may explain in part why the first two generations are
much lighter than the top. If $\Lambda$ is too large, K\"ahler
potential  contributions to fermion masses are small, and
Yukawa interactions in the effective superpotential are
required. In any case, flavor textures above $\Lambda$ need
not resemble those of the standard model.

Decoupling of flavor violation is  a feature of the standard
model, but not of the MSSM. A major advantage of Effective
Supersymmetry is suppression of flavor violation for the first
two generations. Thus the theory is much less contrained  by
FCNC and rare decay limits than the MSSM.

\subsec{Unification}

One of the great successes of the MSSM, the unification of
couplings \nref\unification{ S. Dimopoulos, S. Raby, F. Wilczek,
Phys. Rev. {\bf D24} (1981) 1681.}\refs{\dimgeorg,\unification},
can be easily preserved in
Effective Supersymmetry.  If $\Lambda\ge M_{\rm GUT}$ this
follows trivially, if the only new particles at $\tilde M$ are
the first two generation sparticles. If $\Lambda < M_{\rm
GUT}$ knowledge of the effective theory above $\Lambda$ is
needed to determine whether coupling constants  unify. Even
for low $\Lambda$ the coupling constant unification of the MSSM
can be preserved provided that the accidental approximate
symmetry group $\CA$ above the scale $\tilde M$ contains a
global $SU(5)$ with the standard model gauge group as a
subgroup, and, except for the Higgs doublets, particles come
in approximately degenerate $SU(5)$ multiplets.  As discussed
above such a scenario is desirable since it explains the
absence of the $U(1)_Y$ $D$-term, eq. \mhconst, which would
destabilize the hierarchy.

\subsec{Cosmology}
Gravitino properties depend on the scale $\sqrt F_s$, since
the gravitino eats the goldstino and acquires the mass given
in eq. \mthreehalf. If $\sqrt{F_s}\ltap 10^{10}$~GeV, the
gravitino is the lightest supersymmetric particle.
Cosmological implications of a light gravitino are studied in
\ref\grav{H. Pagels and J.R. Primack, Phys. Rev. Lett. {\bf
48} (1982) 223; T. Moroi, H. Murayama, M. Yamaguchi, Phys.
Lett. {\bf B303} (1993) 289.}. 

With a light top squark and large  CP  violation in soft
supersymmetry breaking terms, the cosmological asymmetry
between baryons and anti-baryons can be generated at the
electroweak phase transition \ref\ewbaryo{A.G. Cohen and  A.E.
Nelson, Phys. Lett. {\bf B297}(1992) 111; S. Myint, Phys.
Lett. {\bf B287} (1992) 325;  P. Huet and  A.E. Nelson, Phys.
Rev. {\bf D53} (1996) 4578; A. Brignole, J.R. Espinosa, M.
Quiros and F. Zwirner, Phys. Lett. {\bf B324} (1994) 181;  M.
Carena, M. Quiros, and C.E.M. Wagner, hep-ph/9603420; J.R.
Espinosa, hep-ph/9604320; J.M. Cline and K. Kainulainen,
hep-ph/9605235.}.

\subsec{Experimental signatures}
The agreement between the standard model and experiment is no
coincidence in our theory. Below $\mt$ the spectrum of our
effective theory is closer to that of the standard model than
the MSSM. Deviations from the standard model in flavor
changing neutral currents, lepton number violation, and
electric dipole moments for the first two generations can be
suppressed below any experimental bounds. If there are large
CP violating phases, and if the top and/or bottom squarks are
lighter than $\sim 550$~GeV, it is possible for the neutron
EDM to be close to current limits \nref\ddlpd{J. Dai, H.
Dykstra, R.G. Leigh, S. Paban and D. Dicus, Phys. Lett. {\bf
B237} (1990) 216, ERRATUM-{\it ibid.} {\bf B242} (1990)
547.}\refs{\dks,\pomarol,\ddlpd}. However   a  detectable top
squark contribution to EDM's requires large mixing between the
left and right handed top squarks, which is very small in many
realizations of Effective Supersymmetry (such as in the
composite example of section~3.1). 

High precision tests of third family couplings and rare $\tau$
and $b$ decay searches could yield evidence for new physics,
since some (and perhaps all) third family sparticles are
light, with masses $\ltap \hbox{1 TeV}$. The weak gauginos and
Higgsinos are also lighter than $\sim 1$ TeV.  Together these
sparticles can contribute to non-standard effects in $b, \tau$,
and top quark physics. In particular CP violation can be
especially interesting at colliders, since there are no longer
any EDM constraints on CP violation in soft supersymmetry
violating terms. 

It is usually assumed that supersymmetry implies a non-minimal
Higgs spectrum; however in Effective Supersymmetry, the
effective theory below $\mt$ may have only a single Higgs
doublet with standard model couplings. In this case the Higgs
mass is nearly the same as in the MSSM with large $\tan\beta$,
large $m_A$, and small left-right squark mixing---\ie\
between $M_Z$ and $\sim 120$ GeV \mssmhiggs.

Currently most experimental search strategies for sparticles,
especially at hadron colliders, rely on the assumption of
squark and slepton degeneracy, and so considerable
modification will be needed to search for Effective
Supersymmetry. For instance most gluino decays will involve
third family quarks and a chargino or neutralino. 

Standard supersymmetry searches  assume that the lightest
supersymmetric particle is a stable neutralino. However, for
$\sqrt{F_s}\ltap 10^6$~GeV the neutralino decay to a gravitino
and either a photon, a $Z$, or a Higgs would take place within
the detector \ref\grav{ M. Dine, A.E. Nelson and  Y. Shirman,
Phys. Rev. {\bf D51} (1995) 1362, S. Dimopoulos, M. Dine, S.
Raby and S. Thomas, Phys. Rev. Lett. {\bf 76} (1996) 3494, S.
Ambrosanio, G.L. Kane, G.D. Kribs, S.P. Martin, and S. Mrenna,
hep-ph/9605398, K.S. Babu, C. Kolda, and F. Wilczek,
hep-ph/9605408, S. Dimopoulos, S. Thomas and  J.D. Wells,  
hep-ph/9604452,
Phys. Rev. D {\sl to appear}. }. 

If the lightest sparticles are found and their interactions
are studied, effective supersymmetry predicts their dimension
4 couplings should nearly agree with standard SUSY, since the
lightest sparticles are necessarily those with the smallest
couplings to the supersymmetry breaking sector. The
dimensionless couplings of the heavier sparticles would
deviate from SUSY predictions by an amount $F_s/\Lambda^2$. 

\newsec{Summary}

It is commonly assumed that the MSSM  with a desert above
$\sim 1$~TeV is the minimal way to reproduce the successes of
the standard model and predict the correct value of
$\sin^2\theta_w$, while maintaining naturalness in the
electroweak symmetry breaking sector. In the MSSM avoiding
FCNC requires very precise squark universality or alignment. 
We have argued that such a scenario is neither minimal nor
necessarily superior. Instead we have advocated that  physics
beyond the $Z$ be constructed from a low energy, effective
field theory perspective. We find that low energy
phenomenology favors a scale $\mt\sim 5$--$20$~TeV for the
first two families of squarks and sleptons, while naturalness
favors a scale below $1$~TeV for the top and left-handed bottom
squarks---implying an intimate connection between the physics
of flavor and the supersymmetry breaking mechanism. New gauge
interactions for the first two families, which are connected
with supersymmetry breaking, can explain such a  sparticle
spectrum while  suppressing B and L violation. Although the
effective theory below $\mt$ is not at all supersymmetric,
quadratically divergent contributions to the Higgs mass
approximately cancel. This framework, which we call Effective
Supersymmetry, has distinctive features that can be tested at
future colliders.

\bigskip
\noindent{\bf ACKNOWLEDGMENTS}

A.C. was supported in part by the DOE under grant
\#DE-FG02-91ER40676. D.K. was supported in part by DOE grant
DOE-ER-40561, and NSF Presidential Young Investigator award
\pyidk. A.N. was supported in part by the DOE under grant
\#DE-FG03-96ER40956. We thank T. Appelquist, S. Dimopoulos,
H. Haber, G. Kane,  A. Pomarol, M. Schmaltz, M. Strassler, 
and S. Thomas for very useful conversations, E. Dudas and P.
Binetruy for sharing an early version of their manuscript, and
the Aspen Center for Physics for its hospitality.

\listrefs

\bye